\documentclass{esub2acm}

\title{Inductive Analysis of the Internet Protocol TLS}
\author{Lawrence C. Paulson, University of Cambridge}

\usepackage{graphicx}
    \let\ts=\thinspace
    \let\Imp=\Longrightarrow

\newcommand\const[1]{\hbox{\sf#1}}

\makeatletter
    \begingroup
    \lccode`\~=`\'
    \lowercase{\endgroup
    \newenvironment{alltt}{%
      \trivlist
      \item\relax
        \if@minipage
        \else
          \vskip\parskip
        \fi
        \leftskip\@totalleftmargin
        \rightskip\z@skip
        \parindent\z@
        \parfillskip\@flushglue
        \parskip\z@skip
        \@@par
        \@tempswafalse
        \def\par{%
          \if@tempswa
            \leavevmode\null\@@par\penalty\interlinepenalty
        \else
          \@tempswatrue
          \ifhmode\@@par\penalty\interlinepenalty\fi
        \fi}
        \obeylines
        \verbatim@font
        \let\org@prime~%
        \@noligs
        \everymath\expandafter{\the\everymath
          \catcode`\'=12 \let~\org@prime}
        \everydisplay\expandafter{\the\everydisplay
          \catcode`\'=12 \let~\org@prime}
        \let\org@dospecials\dospecials
        \g@remfrom@specials{\\}
        \g@remfrom@specials{\{}
        \g@remfrom@specials{\}}
        \let\do\@makeother
        \dospecials
        \let\dospecials\org@dospecials
        \frenchspacing\@vobeyspaces
        \everypar \expandafter{\the\everypar \unpenalty}}
    {\endtrivlist}}
    \def\g@remfrom@specials#1{%
      \def\@new@specials{}
      \def\@remove##1{%
        \ifx##1#1\else
        \g@addto@macro\@new@specials{\do ##1}\fi}
      \let\do\@remove\dospecials
      \let\dospecials\@new@specials
      }

    \def\eqalign#1{\null\,\vcenter{\openup\jot\m@th
      \ialign{\strut\hfil$\displaystyle{##}$&$\displaystyle{{}##}$\hfil
          \crcr#1\crcr}}\,}

\makeatother

    \def\ttbraces{\chardef\{=`\{\chardef\}=`\}\chardef\|=`\\}

    \newenvironment{alltt*}{\begin{alltt}\footnotesize\ttbraces\tt}{\end{alltt}}

    \newenvironment{ttbox}{\begin{quote}\samepage\begin{alltt*}}%
                          {\end{alltt*}\end{quote}}

    \def\lbb{\mathopen{\{\kern-.30em|}}
    \def\rbb{\mathclose{|\kern-.32em\}}}
    \def\comp#1{\lbb#1\rbb}
    \newcommand\INS[1]{\{#1\}\cup}

    \newcommand\Agent{\mathop{\const{Agent}}}
    \newcommand\Nonce{\mathop{\const{Nonce}}}
    \newcommand\Number{\mathop{\const{Number}}}
    \newcommand\Na{\mathit{Na}}
    \newcommand\Nb{\mathit{Nb}}
    \newcommand\Key{\mathop{\const{Key}}}
    \newcommand\Ka{\mathit{Ka}}
    \newcommand\Kb{\mathit{Kb}}
    
    \newcommand\Hash{\mathop{\const{Hash}}\nolimits}
    \newcommand\Crypt{\mathop{\const{Crypt}}}
    \newcommand\parts{\mathop{\const{parts}}}
    \newcommand\analz{\mathop{\const{analz}}}
    \newcommand\synth{\mathop{\const{synth}}}
    \newcommand\bad{\mathop{\const{bad}}}

    \newcommand\NOTES{\mathop{\const{Notes}}}
    \newcommand\Says[3]{\const{Says}\,#1\,#2\,#3}
    \newcommand\Notes[2]{\const{Notes}\,#1\,#2}
    \newcommand\spies[1]{\const{spies}\,#1}
    \newcommand\seespy{\spies{evs}}

    \newcommand\cons{\mathbin{\#}}

    \newcommand\used{\mathop{\const{used}}}
    \newcommand\pubK{\mathop{\const{pubK}}}
    \newcommand\priK{\mathop{\const{priK}}}

\let\rulen=\textit
\newcommand\prm[1]{#1$'$}   
\newcommand\prmm[1]{#1$''$} 

\newcommand\range{\mathop{\const{range}}}
\newcommand\cert{\mathop{\const{certificate}}}
\newcommand\clientK{\mathop{\const{clientK}}}
\newcommand\serverK{\mathop{\const{serverK}}}
\newcommand\sessionK{\mathop{\const{sessionK}}}
\newcommand\PRF{\mathop{\const{PRF}}}
\newcommand\tls{\mathop{\const{tls}}}

\newcommand\PMS{\mathop{\mathit{PMS}}}

\begin{document}

\begin{bottomstuff}
  The research was funded by the U.K.'s Engineering and Physical
     Sciences Research Council, grants GR/K77051 `Authentication
  Logics' and GR/K57381 `Mechanizing Temporal Reasoning.'

  \begin{authinfo}
    \address{Computer Laboratory,
             University of Cambridge,
             Cambridge CB2 3QG, England\\
             email \texttt{lcp@cl.cam.ac.uk}}
  \end{authinfo}
  \permission
\end{bottomstuff}

\begin{abstract}
Internet browsers use security protocols to protect sensitive messages.  An 
inductive analysis of TLS (a descendant of SSL 3.0) has been performed using 
the theorem prover Isabelle.  Proofs are based on higher-order logic and make 
no assumptions concerning beliefs or finiteness.  All the obvious security 
goals can be proved; session resumption appears to be secure even if old 
session keys have been compromised. The proofs suggest minor changes to 
simplify the analysis.

TLS, even at an abstract level, is much more complicated than most protocols
that researchers have verified.  Session keys are negotiated rather than
distributed, and the protocol has many optional parts.  Nevertheless, the
resources needed to verify TLS are modest: six man-weeks of effort and three
minutes of processor time.
\end{abstract}




\markboth{L. C. Paulson}{Inductive Analysis of the Internet Protocol TLS}

\maketitle

\section{Introduction}\label{sec:intro}

Internet commerce requires secure communications.  To order goods, a customer
typically sends credit card details.  To order life insurance, the customer
might have to supply confidential personal data.  Internet users would like to
know that such information is safe from eavesdropping or alteration.

Many Web browsers protect transmissions using the protocol SSL (Secure Sockets
Layer).  The client and server machines exchange nonces and compute session
keys from them.  Version 3.0 of SSL has been designed to correct a flaw of
previous versions, the \emph{cipher-suite rollback attack}, whereby an intruder
could get the parties to adopt a weak cryptosystem~\cite{ws-ssl}.  The latest
version of the protocol is called TLS (Transport Layer
Security)~\cite{tls-1.0}; it closely resembles SSL 3.0.

Is TLS really secure?  My proofs suggest that it is, but one should draw no
conclusions without reading the rest of this paper, which describes how the
protocol was modelled and what properties were proved.  I have analyzed a
much simplified form of TLS; I assume hashing and encryption to be secure.

My abstract version of TLS is simpler than the concrete protocol, but it is
still more complex than the protocols typically verified.  We have
not reached the limit of what can be analyzed formally.

The proofs were conducted using Isabelle/HOL~\cite{paulson-isa-book}, an
interactive theorem prover for higher-order logic.  They use the inductive
method~\cite{paulson-jcs}, which has a simple semantics and treats
infinite-state systems.  Model-checking is not used, so there are no
restrictions on the agent population, numbers of concurrent runs, etc.

The paper gives an overview of TLS (\S\ref{sec:overview-tls}) and of the
inductive method for verifying protocols~(\S\ref{sec:proving}).  It continues
by presenting the Isabelle formalization of TLS (\S\ref{sec:formalizing}) and
outlining some of the properties proved~(\S\ref{sec:props-proved}).  Finally,
the paper discusses related work (\S\ref{sec:related}) and
concludes~(\S\ref{sec:concl}).

\section{Overview of TLS}\label{sec:overview-tls}

A TLS \emph{handshake} involves a \emph{client}, such as a World Wide Web
browser, and a Web \emph{server}.  Below, I refer to the client as~$A$
(`Alice') and the server as~$B$ (`Bob'), as is customary for authentication
protocols, especially since $C$ and $S$ often have dedicated meanings in the
literature.

At the start of a handshake, $A$ contacts~$B$, supplying a session identifier
and nonce.  In response, $B$ sends another nonce and his public-key
certificate (my model omits other possibilities).  Then $A$ generates a
\emph{pre-master-secret}, a 48-byte random string, and sends it to $B$
encrypted with his public key.  $A$ optionally sends a signed message to
authenticate herself.  Now, both parties calculate the
\emph{master-secret}~$M$ from the nonces and the pre-master-secret, using a
secure pseudo-random-number function ($\PRF$). They calculate session keys
and MAC secrets from the nonces and master-secret.  Each session involves a
pair of symmetric keys; $A$ encrypts using one and $B$ encrypts using the
other.  Similarly, $A$ and $B$ protect message integrity using separate MAC
secrets.  Before sending application data, both parties exchange
\textbf{finished} messages to confirm all details of the handshake and to
check that cleartext parts of messages have not been altered.

A full handshake is not always necessary.  At some later time,
$A$ can resume a session by quoting an old session identifier along with a
fresh nonce.  If $B$ is willing to resume the designated session, then he
replies with a fresh nonce.  Both parties compute fresh session keys from
these nonces and the stored master-secret, $M$.  Both sides confirm this
shorter run using \textbf{finished} messages.

TLS is highly complex.  My version leaves out many details for the sake of
simplicity:
\begin{itemize}
\item Record formats, field widths, cryptographic algorithms, etc.\ are
  irrelevant in an abstract analysis.
\item Alert and failure messages are unnecessary because bad sessions can
  simply be abandoned.
\item The \textbf{server key exchange} message allows anonymous sessions
  among other things, but it is not an essential part of the protocol.
\end{itemize}

\begin{figure*}

\begin{center}
\includegraphics[scale=0.6]{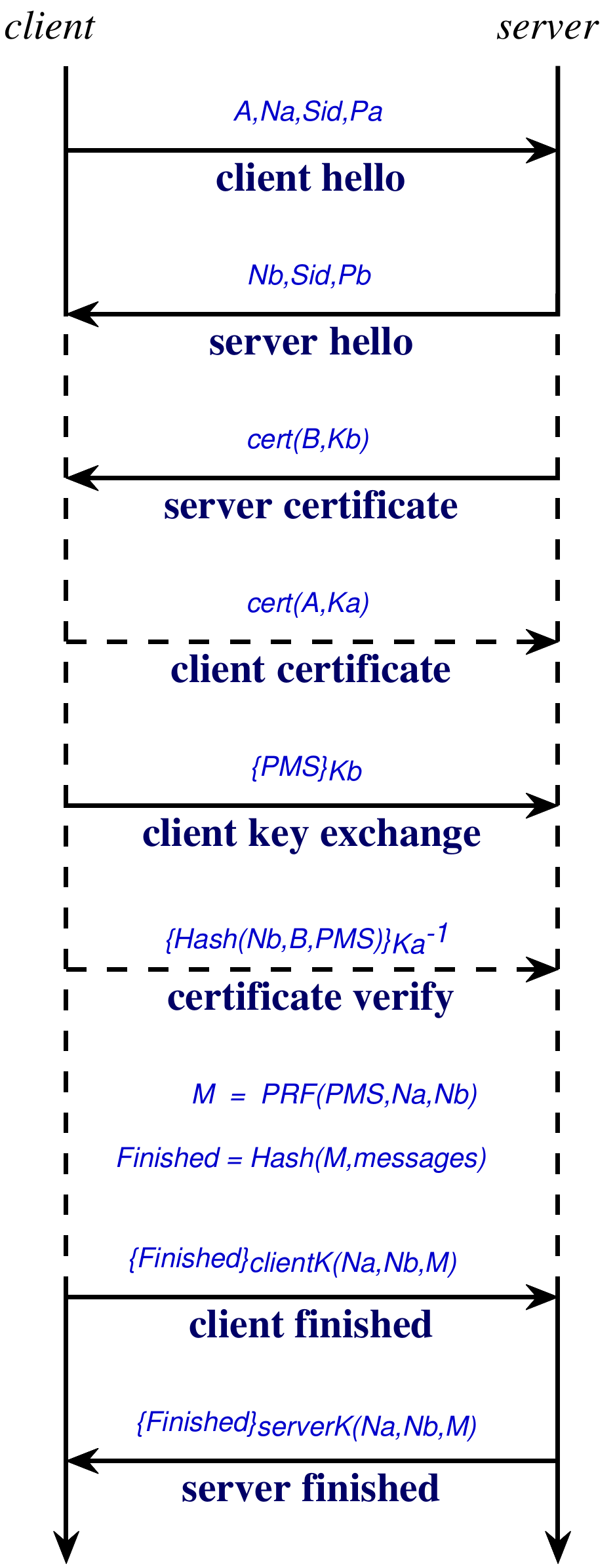}
\end{center}

\caption{The TLS Handshake Protocol as Modelled} \label{fig:tls-eps}
\end{figure*}

Here are the handshake messages in detail, as I model them, along with
comments about their relation to full TLS\@.  Section numbers, such as
tls\S7.3, refer to the TLS specification~\cite{tls-1.0}.  
In Fig.\ts\ref{fig:tls-eps}, dashed lines indicate optional parts.
\[ \hbox{\textbf{client hello}}\quad A\to B : A, \Na, Sid, Pa \]
The items in this message include the nonce $\Na$, called \textbf{client
  random}, and the session identifier $Sid$.  The model makes no
assumptions about the structure of agent names such as $A$ and~$B$.  Item
$Pa$ is
$A$'s set of preferences for encryption and compression; due to export
controls, for example, some clients cannot support certain encryption
methods.  For our purposes, all that matters is that both parties can detect
if $Pa$ has been altered during transmission (tls\S7.4.1.2).%
\footnote{According to the TLS specification, \textbf{client hello} does not
mention the client's name.  But obviously the server needs to know where the
request comes from, and in practice gets this information from
the underlying transport protocol (TCP)\@.  My formalization therefore makes
the sender field explicit.  Note that it is not protected and could be
altered by an intruder.}
\[ \hbox{\textbf{server hello}}\quad B\to A : \Nb, Sid, Pb \]
Agent $B$, in his turn, replies with his nonce~$\Nb$ (\textbf{server
random}).  He repeats the session identifier and returns as~$Pb$ his
cryptographic preferences, selected from~$Pa$.
\[ \hbox{\textbf{server certificate}}\quad B\to A : \cert(B,\Kb) \]
The server's public key, $Kb$, is delivered in a certificate signed by a
trusted third party.  (The TLS proposal (tls\S7.4.2) says it is `generally an
X.509v3 certificate.'  I assume a single certification authority and omit
lifetimes and similar details.)  Making the certificate mandatory and
eliminating the \textbf{server key exchange} message (tls\S7.4.3) simplifies
\textbf{server hello}.  I leave \textbf{certificate request} (tls\S7.4.4)
implicit: $A$ herself decides whether or not to send the optional messages
\textbf{client certificate} and \textbf{certificate verify}.
\[
  \begin{array}{ll}
    \hbox{\textbf{client certificate*}}& A\to B : \cert(A,\Ka) \\[1ex]
    \hbox{\textbf{client key exchange}}& A\to B : \comp{\PMS}_{\Kb} \\[1ex]
    \hbox{\textbf{certificate verify*}}& A\to B 
        : \comp{\Hash\comp{\Nb,B,\PMS}}_{\Ka^{-1}}
  \end{array}
\]
The notation $\comp{X}_{K}$ stands for the message $X$ encrypted or
signed using the key~$K$.  Optional messages are starred (*) above; in
\textbf{certificate verify}, $A$ authenticates herself to~$B$ by signing the
hash of some items relevant to the current session.  The specification
states that all handshake messages should be hashed, but my proofs suggest
that only $\Nb$, $B$ and $\PMS$ are essential.

For simplicity, I do not model the possibility of
arriving at the pre-master-secret via a Diffie-Hellman exchange
(tls\S7.4.7.2).  The proofs therefore can say nothing about this part of the
protocol.
\[
  \begin{array}{ll}
    \hbox{\textbf{client finished}}& A\to B : 
                  \comp{\textit{Finished}\,}_{\clientK(\Na,\Nb,M)} \\[1ex]
    \hbox{\textbf{server finished}}& A\to B : 
                  \comp{\textit{Finished}\,}_{\serverK(\Na,\Nb,M)} 
    \end{array}
\]
Both parties compute the master-secret $M$ from $\PMS$, $\Na$ and~$\Nb$ and
compute $\textit{Finished}$ as the hash of $Sid$, $M$, $\Na$, $Pa$, $A$, $\Nb$,
$Pb$, $B$.  According to the specification (tls\S7.4.9), $M$ should be hashed
with all previous handshake messages using~$\PRF$.  My version hashes
message components rather than messages in order to simplify the inductive
definition; as a consequence, it is vulnerable to an attack in which the spy
intercepts
\textbf{certificate verify}, downgrading the session so that the client
appears to be unauthenticated.

The symmetric key $\clientK(\Na,\Nb,M)$ is intended for client encryption,
while $\serverK(\Na,\Nb,M)$ is for server encryption; each party decrypts
using the other's key (tls\S6.3).  The corresponding MAC secrets are implicit
because my model assumes strong encryption; formally, the only operation that
can be performed on an encrypted message is to decrypt it using the
appropriate key, yielding the original plaintext.  With encryption already
providing an integrity check, there is no need to include MAC secrets in the
model.

Once a party has received the other's \textbf{finished} message and compared
it with her own, she is assured that both sides agree on all critical
parameters, including $M$ and the preferences~$Pa$ and~$Pb$.  Now she may
begin sending confidential data.  The SSL specification~\cite{ssl-3.0}
erroneously states that she can send data immediately after sending her own
\textbf{finished} message, before confirming these parameters; there she takes
a needless risk, since an attacker may have changed the preferences to request
weak encryption.  This is the cipher-suite rollback attack, precisely the one
that the \textbf{finished} messages are intended to prevent.  TLS corrects
this error.

For session resumption, the \textbf{hello} messages are the same.  After
checking that the session identifier is recent enough, the parties exchange
\textbf{finished} messages and start sending application data.
On paper, then, session resumption does not involve any new message types.  
But in the model, four further events are involved.  Each party stores the
session parameters after a successful handshake and looks them up when
resuming a session.

\section{Proving Protocols Using Isabelle}\label{sec:proving}

Isabelle~\cite{paulson-isa-book} is an interactive theorem prover supporting
several formalisms, one of which is higher-order logic (HOL).  Protocols can be
modelled in Isabelle/HOL as inductive definitions.  Isabelle's simplifier and
classical reasoner automate large parts of the proofs.
A security protocol is modelled as the set of traces that could arise when a
population of agents run it.  Among the agents is a spy who controls some
subset of them as well as the network itself.  In constrast to
formalizations intended for model checking, both the population 
and the number of interleaved sessions is unlimited.  This section
summarizes the approach, described in detail elsewhere~\cite{paulson-jcs}.

\subsection{Messages}

Messages are composed of agent names, nonces, keys, etc.:
\[
  \begin{array}{ll}
      \Agent A  & \hbox{identity of an agent}\\
      \Number N & \hbox{guessable number}\\
      \Nonce N  & \hbox{non-guessable number}\\
      \Key K    & \hbox{cryptographic key}\\
      \Hash X   & \hbox{hash of message $X$}\\
      \Crypt K X & \hbox{encryption of $X$ with key $K$}\\
      \comp{X_1,\ldots,X_n} & \hbox{concatenation of messages}
    \end{array}
\]
The notion of \emph{guessable} concerns the spy, who is given the power to
generate any guessable item.  The protocol's \textbf{client
random} and
\textbf{server random} are modelled using $\Nonce$ because they are 28-byte
random values, while
\textbf{session
  identifiers} are modelled using $\Number$ because they may be any strings,
which might be predictable.  TLS sends these items in clear, so whether they
are guessable or not makes little difference to what can be proved.  The
pre-master-secret must be modelled as a nonce; we shall prove no security
properties by assuming it can be guessed.

The model assumes strong encryption.  Hashing is collision-free, and nobody
can recover a message from its hash.  Encrypted messages can neither be read
nor changed without using the corresponding key.  The protocol verifier makes
such assumptions not because they are true but because making them true is the
responsibility of the cryptographer.  Moreover, reasoning about a
cryptosystem such as DES down to the bit level is infeasible.  However, this
is a weakness of the method: certain combinations of protocols and encryption
methods can be vulnerable~\cite{ryan-attack}. 

Three operators are used to express security properties.  Each maps a set~$H$
of messages to another such set.  Typically $H$ is a history of all messages
ever sent, augmented with the spy's initial knowledge of compromised keys.

\begin{itemize}
\item $\parts H$ is the set of message components potentially recoverable
from $H$ (assuming all ciphers could be broken).
\item $\analz H$ is the set of message components recoverable from~$H$
by means of decryption using keys available (recursively) in~$\analz H$.
\item $\synth H$ is the set of messages that could be expressed, starting
from~$H$ and guessable items, using hashing, encryption and concatenation.
\end{itemize}

\subsection{Traces}

A trace is a list of \emph{events} such as $\Says{A}{B}{X}$, meaning `$A$
sends message~$X$ to~$B$,' or $\Notes{A}{X}$, meaning `$A$ stores $X$
internally.'  Each trace is built in reverse order by prefixing (`consing')
events to the front of the list, where \# is the `cons' operator.

The set $\const{bad}$ comprises those agents who are under the spy's
control. 

The function $\const{spies}$ yields the set of messages the spy can see in a
trace: all messages sent across the network and the
internal notes and private keys of the bad agents.
\[ \eqalign{
   \spies{((\Says{A}{B}{X})\cons evs)} &= \INS{X} \spies{evs}\cr
   \spies{((\Notes{A}{X})\cons evs)}   &= 
      \left\{
      \begin{array}{l@{\quad}l} \INS{X} \spies{evs}   &  \hbox{if } A\in\bad\\
                    \spies{evs}               &  \hbox{otherwise}
      \end{array}
      \right.
           }
\]

The set $\used evs$ includes the parts of all messages in the trace, whether
they are visible to other agents or not.  Now $\Na\not\in\used evs$ expresses
that $\Na$ is fresh with respect to the trace~$evs$.
\[ \eqalign{
   \used{((\Says{A}{B}{X})\cons evs)} &= \parts\{X\} \cup \used{evs}\cr
   \used{((\Notes{A}{X})\cons evs)}   &= \parts\{X\} \cup \used{evs}
           }
\]

\section{Formalizing the Protocol in Isabelle}\label{sec:formalizing}

With the inductive method, each protocol step is translated into a rule of an
inductive definition.  A rule's premises describe the conditions under
which the rule may apply, while its conclusion adds new events to the trace.
Each rule allows a protocol step to occur but does not force it to
occur---just as real world machines crash and messages get intercepted.  The
inductive definition has further rules to model intruder actions, etc.

For TLS, the inductive definition comprises fifteen rules, compared with the
usual six or seven for simpler protocols.  The computational cost of proving
a theorem is only linear in the number of rules: proof by induction considers
each rule independently of the others.  But the cost seems to be
exponential in the \emph{complexity} of a rule, for example if there is multiple
encryption.  Combining rules in order to reduce their number is therefore
counterproductive.

\subsection{Basic Constants}

TLS uses both public-key and shared-key encryption.  Each agent $A$ has a
private key $\priK A$ and a public key~$\pubK A$.  The operators $\clientK$
and $\serverK$ create symmetric keys from a triple of nonces.  Modelling the
underlying pseudo-random-number generator causes some complications compared
with the treatment of simple public-key protocols such as
Needham-Schroeder~\cite{paulson-jcs}.

The common properties of $\clientK$ and $\serverK$ are captured in the
function $\sessionK$, which is assumed to be an injective (collision-free)
source of session keys.  In an Isabelle theory file, functions are declared as
constants that have a function type.  Axioms about them can be given using a
\texttt{rules} section.
\begin{ttbox}
datatype role = ClientRole | ServerRole
consts
  sessionK         :: "(nat*nat*nat) * role => key"
  clientK, serverK :: "nat*nat*nat => key"
rules
  inj_sessionK   "inj sessionK" 
  isSym_sessionK "isSymKey (sessionK nonces)"
\end{ttbox}
The enumeration type, $\const{role}$, indicates the use of the session key.
We ensure that $\clientK$ and $\serverK$ have disjoint ranges (no collisions
between the two) by defining
\[ \eqalign{
\clientK X &= \sessionK(X,\,\const{ClientRole}) \cr
\serverK X &= \sessionK(X,\,\const{ServerRole}).
          }
\]
We must also declare the pseudo-random function $\PRF$.
In the real protocol, $\PRF$ has an elaborate definition in terms of the hash
functions MD5 and SHA-1 (see tls\S5).  At the abstract level, we simply assume
$\PRF$ to be injective.  
\begin{ttbox}
consts
  PRF :: "nat*nat*nat => nat"
  tls :: "event list set"
rules
  inj_PRF       "inj PRF"       
\end{ttbox}
We have also declared the constant $\tls$ to be the set of possible traces in
a system running the protocol.  The inductive definition of $\tls$ specifies
it to be the least set of traces that is closed under the rules supplied
below.  A trace belongs to $\tls$ only if it can be generated by finitely many
applications of the rules.  Induction over $\tls$ amounts to considering every
possible way that a trace could have been extended.

\subsection{The Spy}

Figure~\ref{fig:tls-basic} presents the first three rules, two of which are
standard.  Rule \rulen{Nil} allows the empty trace.  Rule \rulen{Fake} says
that the spy may invent messages using past traffic and send them to any other
agent.  A third rule, \rulen{SpyKeys}, augments \rulen{Fake} by letting the
spy use the TLS-specific functions $\sessionK$ and $\PRF$.  In conjunction
with the spy's other powers, it allows him to apply $\sessionK$ and $\PRF$ to
any three nonces previously available to him.  It does not let him invert
these functions, which we assume to be one-way.  We could replace
\rulen{SpyKeys} by defining a TLS version of the function $\synth$; however,
we should then have to rework the underlying theory of messages, which is
common to all protocols.

\begin{figure}[hbtp]
\begin{ttbox}
\textrm{\rulen{Nil}}
[] \(\in\) tls

\textrm{\rulen{Fake}}
[| evs \(\in\) tls;  X \(\in\) synth (analz (spies evs)) |]
\(\Imp\) Says Spy B X  # evs \(\in\) tls


\textrm{\rulen{SpyKeys}}
[| evsSK \(\in\) tls;
   \{|Nonce NA, Nonce NB, Nonce M|\} \(\subseteq\) analz (spies evsSK) |]
\(\Imp\) Notes Spy \{| Nonce (PRF(M,NA,NB)),
                 Key (sessionK((NA,NB,M),role)) |\} # evsSK \(\in\) tls
\end{ttbox}
\caption{Specifying TLS: Basic Rules} \label{fig:tls-basic}
\end{figure}

\subsection{Hello Messages}\label{sec:hello}

Figure~\ref{fig:tls-hello} presents three rules for the \textbf{hello}
messages.  \textbf{Client hello} lets any agent $A$ send the nonce $\Na$,
session identifier~$Sid$ and preferences~$Pa$ to any other agent,~$B$.
\textbf{Server hello} is modelled similarly.  Its precondition is that $B$ has
received a suitable instance of \textbf{Client hello}.

\begin{figure}[hbtp]
\begin{ttbox}
\textrm{\rulen{ClientHello}}
[| evsCH \(\in\) tls;  Nonce NA \(\not\in\) used evsCH;  NA \(\not\in\) range PRF |]
\(\Imp\) Says A B \{|Agent A, Nonce NA, Number SID, Number PA|\}
            # evsCH \(\in\) tls

\textrm{\rulen{ServerHello}}
[| evsSH \(\in\) tls;  Nonce NB \(\not\in\) used evsSH;  NB \(\not\in\) range PRF;
   Says \prm{A} B \{|Agent A, Nonce NA, Number SID, Number PA|\}
    \(\in\) set evsSH |]
\(\Imp\) Says B A \{|Nonce NB, Number SID, Number PB|\} # evsSH \(\in\) tls

\textrm{\rulen{Certificate}}
evsC \(\in\) tls \(\Imp\) Says B A (certificate B (pubK B)) # evsC \(\in\) tls
\end{ttbox}
\caption{Specifying TLS: \textbf{Hello} Messages} \label{fig:tls-hello}
\end{figure}

In \textbf{Client hello}, the assumptions $\Na \not\in\used evsCH$ and
$\Na\not\in \range\PRF$ state that $\Na$ is fresh and distinct from all
possible master-secrets.  The latter assumption precludes the possibility that
$A$ might choose a nonce identical to some master-secret.  (The standard
function $\used$ does not cope with master-secrets because they never appear
in traffic.)  Both assumptions are reasonable because a 28-byte random string
is highly unlikely to clash with any existing nonce or future master-secret.
Still, the condition seems stronger than necessary.  It refers to all
conceivable master-secrets because there is no way of referring to one single
future.  As an alternative, a `no coincidences' condition might be imposed
later in the protocol, but the form it should take is not obvious; if it
is wrong, it might exclude realistic attacks.
  
The \rulen{Certificate} rule handles both \textbf{server certificate} and
\textbf{client certificate}.  It is more liberal than real TLS, for any agent
may send his public-key certificate to any other agent.  A certificate is
represented by an (agent, key) pair signed by the authentication
server.  Freshness of certificates and other details are not modelled.
\begin{ttbox}
constdefs certificate :: "[agent,key] => msg"
         "certificate A KA == Crypt(priK Server)\{|Agent A, Key KA|\}"
\end{ttbox}

\subsection{Client Messages}\label{sec:client-msg}

The next two rules concern \textbf{client key exchange} and \textbf{certificate
  verify} (Fig.\ts\ref{fig:tls-client-key}).  Rule \rulen{ClientKeyExch}
chooses a $\PMS$ that is fresh and differs from all master-secrets, like the
nonces in the \textbf{hello} messages.  It requires \textbf{server
  certificate} to have been received.  No agent is allowed to know the true
sender of a message, so \rulen{ClientKeyExch} might deliver the $\PMS$ to the
wrong agent.  Similarly, \rulen{CertVerify} might use the $\Nb$ value from the
wrong instance of \textbf{server hello}.  Security is not compromised because
the run will fail in the \textbf{finished} messages.

\begin{figure}[hbtp]
\begin{ttbox}
\textrm{\rulen{ClientKeyExch}}
[| evsCX \(\in\) tls;  Nonce PMS \(\not\in\) used evsCX;  PMS \(\not\in\) range PRF;
   Says \prm{B} A (certificate B KB) \(\in\) set evsCX |]
\(\Imp\) Says A B (Crypt KB (Nonce PMS))
      # Notes A \{|Agent B, Nonce PMS|\}
      # evsCX \(\in\) tls

\textrm{\rulen{CertVerify}}
[| evsCV \(\in\) tls;  
   Says \prm{B} A \{|Nonce NB, Number SID, Number PB|\} \(\in\) set evsCV;
   Notes A \{|Agent B, Nonce PMS|\} \(\in\) set evsCV |]
\(\Imp\) Says A B (Crypt (priK A) (Hash\{|Nonce NB, Agent B, Nonce PMS|\}))
              # evsCV \(\in\) tls
\end{ttbox}
\caption{\textbf{Client key exchange} and \textbf{certificate verify}} 
\label{fig:tls-client-key}
\end{figure}

\rulen{ClientKeyExch} not only sends the encrypted $\PMS$ to~$B$ but also
stores it internally using the event $\Notes{A}{\comp{B,\PMS}}$.  Other rules
model $A$'s referring to this note.  For instance, \rulen{CertVerify} states
that if $A$ chose $\PMS$ for~$B$ and has received a \textbf{server hello}
message, then she may send \textbf{certificate verify}.  

In my initial work on TLS, I modelled $A$'s knowledge by referring to the
event of her sending $\comp{\PMS}_{Kb}$ to~$B$.  However, this approach did
not correctly model the sender's knowledge: the spy can intercept and send the
ciphertext $\comp{\PMS}_{Kb}$ without knowing $\PMS$.  (The approach does work
for shared-key encryption.  A ciphertext such as $\comp{\PMS}_{Kab}$
identifies the agents who know the plaintext, namely $A$ and~$B$.)  I
discovered this anomaly when a proof failed.  The final proof state indicated
that the spy could gain the ability to send \textbf{client finished} merely by
replaying $A$'s message $\comp{\PMS}_{Kb}$.

Anomalies like this one can creep into any formalization.  The worst are those
that make a theorem hold vacuously, for example by mis-stating a precondition.
There is no remedy but constant vigilance, noticing when a result is too good
to be true or is proved too easily.  We must also check that the assumptions
built into the model, such as strong encryption, reasonably match the
protocol's operating environment.

\subsection{Finished Messages}\label{sec:finished}

Next come the \textbf{finished} messages (Fig.\ts\ref{fig:tls-finished}).
\rulen{ClientFinished} states that if $A$ has sent \textbf{client hello} and
has received a plausible instance of \textbf{server hello} and has chosen a
$\PMS$ for~$B$, then she can calculate the master-secret and send a
\textbf{finished} message using her \textbf{client write key}.
\rulen{ServerFinished} is analogous and may occur if $B$ has received a
\textbf{client hello}, sent a \textbf{server hello}, and received a
\textbf{client key exchange} message.

\begin{figure}[hbtp]
\begin{ttbox}
\textrm{\rulen{ClientFinished}}
[| evsCF \(\in\) tls;  
   Says A  B \{|Agent A, Nonce NA, Number SID, Number PA|\} \(\in\){\ts}set evsCF;
   Says \prm{B} A \{|Nonce NB, Number SID, Number PB|\} \(\in\) set evsCF;
   Notes A \{|Agent B, Nonce PMS|\} \(\in\) set evsCF;
   M = PRF(PMS,NA,NB) |]
\(\Imp\) Says A B (Crypt (clientK(NA,NB,M))
              (Hash\{|Number SID, Nonce M,
                     Nonce NA, Number PA, Agent A, 
                     Nonce NB, Number PB, Agent B|\}))
    # evsCF \(\in\) tls

\textrm{\rulen{ServerFinished}}
[| evsSF \(\in\) tls;
   Says \prm{A} B \{|Agent A, Nonce NA, Number SID, Number PA|\} \(\in\){\ts}set evsSF;
   Says B  A \{|Nonce NB, Number SID, Number PB|\} \(\in\) set evsSF;
   Says \prmm{A} B (Crypt (pubK B) (Nonce PMS)) \(\in\) set evsSF;
   M = PRF(PMS,NA,NB) |]
\(\Imp\) Says B A (Crypt (serverK(NA,NB,M))
              (Hash\{|Number SID, Nonce M,
                     Nonce NA, Number PA, Agent A, 
                     Nonce NB, Number PB, Agent B|\}))
    # evsSF \(\in\) tls
\end{ttbox}
\caption{\textbf{Finished} messages}  \label{fig:tls-finished}
\end{figure}

\subsection{Session Resumption}
\label{sec:resumption}

That covers all the protocol messages, but the specification is not complete.
Next come two rules to model agents' confirmation of a session
(Fig.\ts\ref{fig:tls-accepts}).  Each agent, after sending its finished
message and receiving a matching finished message apparently from its peer,
records the session parameters to allow resumption.  Next come two rules for
session resumption (Fig.\ts\ref{fig:tls-resume}).  Like \rulen{ClientFinished}
and \rulen{ServerFinished}, they refer to two previous hello messages.  But
instead of calculating the master-secret from a $\PMS$ just sent, they use the
master-secret stored by \rulen{ClientAccepts} or \rulen{ServerAccepts} with
the same session identifier.  They calculate new session keys using the fresh
nonces.

\begin{figure}[hbtp]
\begin{ttbox}
\textrm{\rulen{ClientAccepts}}
[| evsCA \(\in\) tls;
   Notes A \{|Agent B, Nonce PMS|\} \(\in\) set evsCA;
   M = PRF(PMS,NA,NB);  
   X = Hash\{|Number SID, Nonce M,
             Nonce NA, Number PA, Agent A, 
             Nonce NB, Number PB, Agent B|\};
   Says A  B (Crypt (clientK(NA,NB,M)) X) \(\in\) set evsCA;
   Says \prm{B} A (Crypt (serverK(NA,NB,M)) X) \(\in\) set evsCA |]
\(\Imp\) Notes A \{|Number SID, Agent A, Agent B, Nonce M|\} # evsCA \(\in\) tls

\textrm{\rulen{ServerAccepts}}
[| evsSA \(\in\) tls;  A \(\not=\) B;
   Says \prmm{A} B (Crypt (pubK B) (Nonce PMS)) \(\in\) set evsSA;
   M = PRF(PMS,NA,NB);  
   X = Hash\{|Number SID, Nonce M,
             Nonce NA, Number PA, Agent A, 
             Nonce NB, Number PB, Agent B|\};
   Says B  A (Crypt (serverK(NA,NB,M)) X) \(\in\) set evsSA;
   Says \prm{A} B (Crypt (clientK(NA,NB,M)) X) \(\in\) set evsSA |]
\(\Imp\) Notes B \{|Number SID, Agent A, Agent B, Nonce M|\} # evsSA \(\in\) tls
\end{ttbox}
\caption{Agent acceptance events}  \label{fig:tls-accepts}
\end{figure}

The references to $\PMS$ in the \emph{Accepts} rules appear to contradict the
protocol specification (tls\S8.1): `the pre-master-secret should be deleted
from memory once the master-secret has been computed.'  The purpose of those
references is to restrict the rules to agents who actually know the secrets,
as opposed to a spy who merely has replayed messages (recall the comment at
the end of~\S\ref{sec:client-msg}).  They can probably be replaced by
references to the master-secret, which the agents keep in memory.  We would
have to add further events to the inductive definition.  Complicating the
model in this way brings no benefits: the loss of either secret is equally
catastrophic.

\begin{figure}[hbtp]
\begin{ttbox}
\textrm{\rulen{ClientResume}}
[| evsCR \(\in\) tls;  
   Says A  B \{|Agent A, Nonce NA, Number SID, Number PA|\} \(\in\){\ts}set evsCR;
   Says \prm{B} A \{|Nonce NB, Number SID, Number PB|\} \(\in\) set evsCR;
   Notes A \{|Number SID, Agent A, Agent B, Nonce M|\} \(\in\) set evsCR |]
\(\Imp\) Says A B (Crypt (clientK(NA,NB,M))
              (Hash\{|Number SID, Nonce M,
                     Nonce NA, Number PA, Agent A, 
                     Nonce NB, Number PB, Agent B|\}))
    # evsCR \(\in\) tls

\textrm{\rulen{ServerResume}}
[| evsSR \(\in\) tls;
   Says \prm{A} B \{|Agent A, Nonce NA, Number SID, Number PA|\} \(\in\){\ts}set evsSR;
   Says B  A \{|Nonce NB, Number SID, Number PB|\} \(\in\) set evsSR;  
   Notes B \{|Number SID, Agent A, Agent B, Nonce M|\} \(\in\) set evsSR |]
\(\Imp\) Says B A (Crypt (serverK(NA,NB,M))
              (Hash\{|Number SID, Nonce M,
                     Nonce NA, Number PA, Agent A, 
                     Nonce NB, Number PB, Agent B|\})) # evsSR
     \(\in\) tls
\end{ttbox}
\caption{Agent resumption events}  \label{fig:tls-resume}
\end{figure}
Four further rules (omitted here) model agents' confirmation of a session and
a subsequent session resumption.

\subsection{Security Breaches}\label{sec:oops}

The final rule, \rulen{Oops}, models security breaches.  Any session key, if
used, may end up in the hands of the spy.  Session resumption turns out to be
safe even if the spy has obtained session keys from earlier sessions.
\begin{ttbox}
\textrm{\rulen{Oops}}
[| evso \(\in\) tls;  
   Says A B (Crypt (sessionK((NA,NB,M),role)) X) \(\in\) set evso |]
\(\Imp\) Says A Spy (Key (sessionK((NA,NB,M),role))) # evso \(\in\) tls
\end{ttbox}
Other security breaches could be modelled.  The pre-master-secret might be
lost to a cryptanalytic attack against the \textbf{client key exchange}
message, and \citeN[\S4.7]{ws-ssl} suggest a strategy for discovering the
master-secret.  Loss of the $\PMS$ would compromise the entire session; it is
hard to see what security goal could still be proved (in contrast, loss of a
session key compromises that key alone).  Recall that the spy already controls
the network and an unknown number of agents.

The protocol, as modelled, is too liberal and is highly nondeterministic.  As
in TLS itself, some messages are optional (\textbf{client certificate},
\textbf{certificate verify}).  Either client or server may be the first to
commit to a session or to send a \textbf{finished} message.  One party might
attempt session resumption while the other runs the full protocol.  Nothing in
the rules above stops anyone from responding to any message repeatedly.
Anybody can send a certificate to anyone else at any time.

Such nondeterminism is unacceptable in a real protocol, but it simplifies the
model.  Constraining a rule to follow some other rule or to apply at most once
requires additional preconditions.  A simpler model generally allows simpler
proofs.  Safety theorems proved under a permissive regime will continue to
hold under a strict one.

\section{Properties Proved of TLS}\label{sec:props-proved}

One difficulty in protocol verification is knowing what to prove.  Protocol
goals are usually stated informally.  The TLS memo states `three basic
properties' (tls\S1):
\begin{enumerate}
\item `The peer's identity can be authenticated using \ldots{}
       public key cryptography'

\item `The negotiated secret is unavailable to eavesdroppers, and for any
authenticated connection the secret cannot be obtained, even by an attacker who
can place himself in the middle of the connection'

\item `no attacker can modify the negotiation communication without being
detected by the parties'
\end{enumerate}

Authentication can mean many things~\cite{gollmann-what}; it is a pity that
the memo does not go into more detail.  I have taken `authenticated
connection' to mean one in which both parties use their private keys.  My
model allows $A$ to be unauthenticated, since \textbf{certificate verify} is
optional.  However, $B$ must be authenticated: the model does not support
Diffie-Hellman, so $\Kb^{-1}$ must be used to decrypt \textbf{client key
  exchange}.  Against an active intruder, an unauthenticated connection is
vulnerable to the usual man-in-the-middle attack.  Since the model does not
support unauthenticated connections, I cannot investigate whether they are
secure against passive eavesdroppers.

Some of the results discussed below relate to authentication.  A pair of
honest agents can establish the master-secret securely and use it to generate
uncompromised session keys.  Session resumption is secure even if previous
session keys from that session have been compromised.

\subsection{Basic Lemmas}

In the inductive method, results are of three sorts: possibility properties,
regularity lemmas and secrecy theorems. Possibility properties merely exercise
all the rules to check that the model protocol can run.  For a simple protocol,
one possibility property suffices to show that message formats are compatible. 
For TLS, I proved four properties to check various paths through the main
protocol, the \textbf{client verify} message, and session resumption.

Regularity lemmas assert properties that hold of all traffic.  For example, no
protocol step compromises a private key.  From our specification of TLS, it is
easy to prove that all certificates are valid.  (This property is overly
strong, but adding false certificates seems pointless: $B$ might be under the
spy's control anyway.)  If $\cert(B,K)$ appears in traffic, then $K$ really is
$B$'s public key:
\begin{ttbox}
[| certificate B KB \(\in\) parts(spies evs);  evs \(\in\) tls |] \(\Imp\) pubK B = KB
\end{ttbox}
The set $\parts(\spies
evs)$ includes the components of all messages that have been sent; in the
inductive method, regularity lemmas often mention this set.  Sometimes the
lemmas merely say that events of a particular form never occur.

Many regularity lemmas are technical.  Here are two typical ones.  If a
master-secret has appeared in traffic, then so has the underlying
pre-master-secret.  Only the spy might send such a message.
\begin{ttbox}
[| Nonce (PRF (PMS,NA,NB)) \(\in\) parts(spies evs);  evs \(\in\) tls |]  
\(\Imp\) Nonce PMS \(\in\) parts(spies evs)
\end{ttbox}
If a pre-master-secret is fresh, then no session key derived from it can either
have been transmitted or used to encrypt.%
\footnote{The two properties must be proved in
mutual induction because of interactions between the Fake and Oops rules.}
\begin{ttbox}
[| Nonce PMS \(\not\in\) parts(spies evs);
   K = sessionK((Na, Nb, PRF(PMS,NA,NB)), role);  
   evs \(\in\) tls |] 
\(\Imp\) Key K \(\not\in\) parts(spies evs) & (\(\forall\) Y. Crypt K Y \(\not\in\) parts(spies evs))
\end{ttbox}
Client authentication, one of the protocol's goals, is easily proved.  If
\textbf{certificate verify} has been sent, apparently by~$A$, then it really
has been sent by~$A$ provided $A$ is uncompromised (not controlled by the
spy).  Moreover, $A$ has chosen the pre-master-secret that is hashed in
\textbf{certificate verify}.
\begin{ttbox}
[| X \(\in\) parts(spies evs);  X = Crypt KA\(\sp{-1}\) (Hash\{|nb, Agent B, pms|\}); 
   certificate A KA \(\in\) parts(spies evs);             
   evs \(\in\) tls;  A \(\not\in\) bad |] 
\(\Imp\) Says A B X \(\in\) set evs
\end{ttbox}

\subsection{Secrecy Goals}\label{sec:secrecy}

Other goals of the protocol relate to secrecy: certain items are available to
some agents but not to others.  They are usually the
hardest properties to establish.  With the inductive method, they seem always
to require, as a lemma, some form of \emph{session key compromise theorem}.
This theorem imposes limits on the message components that can become
compromised by the loss of a session key.  Typically we require that these
components contain no session keys, but for TLS, they must contain no nonces.
Nonces are of critical importance because one of them is the
pre-master-secret.

The theorem seems obvious.  No honest agent encrypts nonces using session
keys, and the spy can only send nonces that have already been compromised.
However, its proof takes over seven seconds to run.  Like other secrecy
proofs, it involves a large, though automatic, case analysis.
\begin{ttbox}
evs \(\in\) tls \(\Imp\)       
Nonce N \(\in\) analz (insert (Key (sessionK z)) (spies evs)) = 
(Nonce N \(\in\) analz (spies evs))
\end{ttbox}
Note that $\const{insert}\,x\,A$ denotes $\{x\}\cup A$.  The set
$\analz(\seespy)$ includes all message components available to the spy, and
likewise $\analz({\{K\}} \cup \seespy)$ includes all message components that
the spy could get with the help of key~$K$.  The theorem states that session
keys do not help the spy to learn new nonces.

Other secrecy proofs follow easily from the session key compromise theorem,
using induction and simplification.  Provided $A$ and $B$ are honest, the
client's session key will be secure unless $A$ herself gives it to the spy,
using Oops.  
\begin{ttbox}
[| Notes A \{|Agent B, Nonce PMS|\} \(\in\) set evs;
   Says A Spy (Key (clientK(NA,NB,PRF(PMS,NA,NB)))) \(\not\in\) set evs; 
   A \(\not\in\) bad;  B \(\not\in\) bad;  evs \(\in\) tls |]   
\(\Imp\) Key (clientK(NA,NB,PRF(PMS,NA,NB))) \(\not\in\) parts(spies evs)
\end{ttbox}
An analogous theorem holds for the server's session key.  However, the server
cannot check the $\NOTES$ assumption; see~\S\ref{sec:trust-client}.
\begin{ttbox}
[| Notes A \{|Agent B, Nonce PMS|\} \(\in\) set evs;
   Says B Spy (Key (serverK(NA,NB,PRF(PMS,NA,NB)))) \(\not\in\) set evs; 
   A \(\not\in\) bad;  B \(\not\in\) bad;  evs \(\in\) tls |]   
\(\Imp\) Key (serverK(NA,NB,PRF(PMS,NA,NB))) \(\not\in\) parts(spies evs)
\end{ttbox}
If $A$ sends the \textbf{client key exchange} message to~$B$, and both agents
are uncompromised, then the pre-master-secret and master-secret will stay
secret.
\begin{ttbox}
[| Notes A \{|Agent B, Nonce PMS|\} \(\in\) set evs;
   evs \(\in\) tls;  A \(\not\in\) bad;  B \(\not\in\) bad |]           
\(\Imp\) Nonce PMS \(\not\in\) analz(spies evs)
\end{ttbox}

\begin{ttbox}
[| Notes A \{|Agent B, Nonce PMS|\} \(\in\) set evs;
   evs \(\in\) tls;  A \(\not\in\) bad;  B \(\not\in\) bad |]           
\(\Imp\) Nonce (PRF(PMS,NA,NB)) \(\not\in\) analz(spies evs)
\end{ttbox}

\subsection{\textbf{Finished} Messages}

Other important protocol goals concern authenticity of the \textbf{finished}
message.  If each party can know that the \textbf{finished} message just
received indeed came from the expected agent, then they can compare the
message components to confirm that no tampering has occurred.  These
components include the cryptographic preferences, which an intruder might like
to downgrade.  Naturally, the guarantees are conditional on both agents' being
uncompromised.

\subsubsection{Client's guarantee}

The client's guarantee has several preconditions.  The client, $A$, has chosen
a pre-master-secret $\PMS$ for~$B$.  The traffic contains a \textbf{finished}
message encrypted with a \textbf{server write key} derived from
$\PMS$.  The server, $B$, has not given that session key to the spy (via
\rulen{Oops}).  The guarantee then states that $B$ himself has sent that
message, and to~$A$.
\begin{ttbox}
[| X = Crypt (serverK(Na,Nb,M))                  
             (Hash\{|Number SID, Nonce M,             
                    Nonce Na, Number PA, Agent A,    
                    Nonce Nb, Number PB, Agent B|\}); 
   M = PRF(PMS,NA,NB);                           
   X \(\in\) parts(spies evs);                        
   Notes A \{|Agent B, Nonce PMS|\} \(\in\) set evs;     
   Says B Spy (Key (serverK(Na,Nb,M))) \(\not\in\) set evs; 
   evs \(\in\) tls;  A \(\not\in\) bad;  B \(\not\in\) bad |]          
\(\Imp\) Says B A X \(\in\) set evs
\end{ttbox}
One of the preconditions may seem to be too liberal.  The guarantee applies to
any occurrence of the \textbf{finished} message in traffic, but it is needed
only when $A$ has received that message.  The form shown, expressed using
$\parts(\seespy)$, streamlines the proof; in particular, it copes with the
spy's replaying a \textbf{finished} message concatenated with other material.
It is well known that proof by induction can require generalizing the theorem
statement.

\subsubsection{Server's guarantee}\label{sec:trust-client}

The server's guarantee is slightly different.  If any message has been
encrypted with a \textbf{client write key} derived from a given $\PMS$---which
we \emph{assume} to have come from~$A$---and if $A$ has not given that session
key to the spy, then $A$ herself sent that message, and to~$B$.
\begin{ttbox}
[| M = PRF(PMS,NA,NB);                           
   Crypt (clientK(Na,Nb,M)) Y \(\in\) parts(spies evs);  
   Notes A \{|Agent B, Nonce PMS|\} \(\in\) set evs;        
   Says A Spy (Key(clientK(Na,Nb,M))) \(\not\in\) set evs;  
   evs \(\in\) tls;  A \(\not\in\) bad;  B \(\not\in\) bad |]
\(\Imp\) Says A B (Crypt (clientK(Na,Nb,M)) Y) \(\in\) set evs
\end{ttbox}
The assumption (involving $\NOTES$) that $A$ chose the $\PMS$ is essential.
If the client has not authenticated herself, then $B$ knows nothing about her
true identity and must trust that she is indeed~$A$.  By sending
\textbf{certificate verify}, the client can discharge the $\NOTES$ assumption:
\begin{ttbox}
[| Crypt KA\(\sp{-1}\) (Hash\{|nb, Agent B, Nonce PMS|\}) \(\in\) parts(spies evs);                             
   certificate A KA \(\in\) parts(spies evs);              
   evs \(\in\) tls;  A \(\not\in\) bad |]                            
\(\Imp\) Notes A \{|Agent B, Nonce PMS|\} \(\in\) set evs
\end{ttbox}
$B$'s guarantee does not even require his inspecting the
\textbf{finished} message.  The very use of \texttt{clientK(Na,Nb,M)} is proof
that the communication is from $A$ to~$B$. If we consider the analogous
property for~$A$, we find that using \texttt{serverK(Na,Nb,M)} only guarantees
that the sender is~$B$; in the absence of \textbf{certificate verify}, $B$ has
no evidence that the $\PMS$ came from~$A$.  If he sends \textbf{server
  finished} to somebody else then the session will fail, so there is no
security breach.  

Still, changing \textbf{client key exchange} to include $A$'s identity,
\[ A\to B : \comp{{ A}, \PMS}_{\Kb}, \]
would slightly strengthen the protocol and simplify the analysis.  At present,
the proof scripts include theorems for $A$'s association of $\PMS$ with~$B$,
and weaker theorems for $B$'s knowledge of~$\PMS$.  With the suggested change,
the weaker theorems could probably be discarded.

The guarantees for \textbf{finished} messages apply to session resumption as
well as to full handshakes.  The inductive proofs cover all the rules that
make up the definition of the constant~$\tls$, including those that model
resumption.

\subsection{Security Breaches}

The Oops rule makes the model much more realistic.  It allows session keys to
be lost to determine whether the protocol is robust: one security
breach should not lead to a cascade of others.  Sometimes a theorem holds only
if certain Oops events are excluded, but Oops conditions should be weak.
For the \textbf{finished} guarantees, the conditions they impose on Oops
events are as weak as could be hoped for: that the very session key in
question has not been lost by the only agent expected to use that key for
encryption.

\section{Related Work}\label{sec:related}

\citeN{ws-ssl} analyze SSL~3.0 in detail.  Much of their discussion concerns
cryptanalytic attacks.  Attempting repeated session resumptions causes the
hashing of large amounts of known plaintext with the master-secret, which
could lead to a way of revealing it (\S4.7).  They also report an attack
against the Diffie-Hellman key-exchange messages, which my model omits
(\S4.4).  Another attack involves deleting the \textbf{change cipher spec}
message that (in a draft version of SSL 3.0) may optionally be sent before the
\textbf{finished} message.  TLS makes \textbf{change cipher spec} mandatory,
and my model regards it as implicit in the \textbf{finished} exchange.

Wagner and Schneier's analysis appears not to use any formal tools.  
Their form
of scrutiny, particularly concerning attacks against the underlying
cryptosystems, will remain an essential complement to proving protocols at the
abstract level.    

In his PhD thesis, \citeN{dietrich-ssl} analyses SSL~3.0 using
the belief logic NCP (Non-monotonic Cryptographic Protocols).  NCP allows
beliefs to be deleted; in the case of SSL, a session identifier is forgotten
if the session fails.  (In my formalization, session identifiers are not
recorded until the initial session reaches a successful exchange of
\textbf{finished} messages.  Once recorded, they persist forever.)  Recall
that SSL allows both authenticated and unauthenticated sessions; Dietrich
considers the latter and shows them to be secure against a passive
eavesdropper.  Although NCP is a formal logic, Dietrich appears to have
generated his lengthy derivations by hand.

\citeN{mitchell-ssl} apply model checking to a number of simple
protocols derived from SSL~3.0.  Most of the protocols are badly flawed (no
nonces, for example) and the model checker finds many attacks. 
The final
protocol still omits much of the detail of TLS, such as the distinction between
the pre-master-secret and the other secrets computed from it.  An eight-hour
execution found no attacks against the protocol in a system comprising
two clients and one server.

\section{Conclusions}\label{sec:concl}

The inductive method has many advantages.  Its semantic framework, based
on the actions agents can perform, has few of the peculiarities of belief
logics.  Proofs impose no limits on the number of
simultaneous or resumed sessions.  Isabelle's automatic tools allow the proofs
to be generated with a moderate effort, and they run fast.  The full TLS proof
script runs in 150 seconds on a 300Mhz Pentium.

I obtained the abstract message exchange given in \S\ref{sec:overview-tls} by
reverse engineering the TLS specification.  This process took about two weeks,
one-third of the time spent on this verification.  SSL must have originated in
such a message exchange, but I could not find one in the literature.  If
security protocols are to be trusted, their design process must be
transparent.  The underlying abstract protocol should be exposed to public
scrutiny.  The concrete protocol should be presented as a faithful realization
of the abstract one.  Designers should distinguish between attacks against the
abstract message exchange and those against the concrete protocol.

All the expected security goals were proved: no attacks were found.  This
unexciting outcome might be expected in a protocol already so thoroughly
examined.  No unusual lines of reasoning were required, unlike the proofs of
the Yahalom protocol~\cite{paulson-yahalom} and
Kerberos~IV~\cite{bella-kerberos4}; we may infer that TLS is well-designed.
The proofs did yield some insights into TLS, such as the possibility of
strengthening \textbf{client key exchange} by including $A$'s identity
(\S\ref{sec:props-proved}).  The main interest of this work lies in the
modelling of TLS, especially its use of pseudo-random number generators.

The protocol takes the \emph{explicitness
  principle} of~\citeN{abadi-prudent} to an extreme.  In several places, it
requires computing the hash of `all preceding handshake messages.'  There is
obviously much redundancy, and the requirement is ambiguous too; the
specification is sprinkled with remarks that certain routine messages or
components should not be hashed.  One such message, \textbf{change cipher
  spec}, was thereby omitted and later was found to be
essential~\cite{ws-ssl}.  I suggest, then, that hashes should be computed
not over everything but over selected items that the protocol designer
requires to be confirmed.  An inductive analysis can help in selecting the
critical message components.  The TLS security analysis (tls\S F.1.1.2) states
that the critical components of the hash in \textbf{certificate verify} are
the server's name and nonce, but my proofs suggest that the pre-master-secret
is also necessary.

Once session keys have been established, the parties have a secure channel
upon which they must run a reliable communication protocol.  Abadi tells me
that the TLS \emph{application data protocol} should also be examined, since
this part of SSL once contained errors.  I have considered only the TLS
\emph{handshake protocol}, where session keys are negotiated.  Ideally, the
application data protocol should be verified separately, assuming an
unreliable medium rather than an enemy.  My proofs assume that application
data does not contain secrets associated with TLS sessions, such as keys and
master-secrets; if it does, then one security breach could lead to many
others.

Previous verification efforts have largely focussed on small protocols of
academic interest.  It is now clear that realistic protocols can be analyzed
too, almost as a matter of routine.  For protocols intended for critical
applications, such an analysis should be required as part of the certification
process.

\begin{acks}
  Mart\'\i n Abadi introduced me to TLS and identified related work.  James
  Margetson pointed out simplifications to the model.  The referees and
  Clemens Ballarin made useful comments.
\end{acks}

\bgroup
\bibliographystyle{esub2acm}

\begin{thebibliography}{}

\bibitem[\protect\citeauthoryear{Abadi and Needham}{Abadi and
  Needham}{1996}]{abadi-prudent}
\bibsc{Abadi, M. and Needham, R.} \bibyear{1996}.
\newblock Prudent engineering practice for cryptographic protocols.
\newblock \bibemphic{IEEE Trans. Softw. Eng.}~\bibemph{22},~1 (Jan.), 6--15.

\bibitem[\protect\citeauthoryear{Bella and Paulson}{Bella and
  Paulson}{1998}]{bella-kerberos4}
\bibsc{Bella, G. and Paulson, L.~C.} \bibyear{1998}.
\newblock {Kerberos} version {IV}: Inductive analysis of the secrecy goals.
\newblock In \bibsc{J.-J. Quisquater, Y.~Deswarte, C.~Meadows, and D.~Gollmann}
  Eds., \bibemphic{Computer Security --- {ESORICS} 98}, LNCS 1485 (1998), pp.\
  361--375. Springer.

\bibitem[\protect\citeauthoryear{Dierks and Allen}{Dierks and
  Allen}{1999}]{tls-1.0}
\bibsc{Dierks, T. and Allen, C.} \bibyear{1999}.
\newblock The {TLS} protocol: Version 1.0.
\newblock Request for Comments: 2246, on the Internet at
  \verb|ftp://ftp.isi.edu/in-notes/rfc2246.txt|.

\bibitem[\protect\citeauthoryear{Dietrich}{Dietrich}{1997}]{dietrich-ssl}
\bibsc{Dietrich, S.} \bibyear{1997}.
\newblock \bibemph{A Formal Analysis of the Secure Sockets Layer Protocol}.
\newblock Ph.\ D. thesis, Adelphi University, Garden City, New York.
\newblock Department of Mathematics and Computer Science.

\bibitem[\protect\citeauthoryear{Freier, Karlton, and Kocher}{Freier
  et~al.}{1996}]{ssl-3.0}
\bibsc{Freier, A.~O., Karlton, P., and Kocher, P.~C.} \bibyear{1996}.
\newblock The {SSL} protocol version 3.0.
\newblock Internet-draft \verb|draft-freier-ssl-version3-02.txt|.

\bibitem[\protect\citeauthoryear{Gollmann}{Gollmann}{1996}]{gollmann-what}
\bibsc{Gollmann, D.} \bibyear{1996}.
\newblock What do we mean by entity authentication?
\newblock In \bibemphic{Symposium on Security and Privacy} (1996), pp.\
  46--54. IEEE Computer Society.

\bibitem[\protect\citeauthoryear{Mitchell, Shmatikov, and Stern}{Mitchell
  et~al.}{1997}]{mitchell-ssl}
\bibsc{Mitchell, J.~C., Shmatikov, V., and Stern, U.} \bibyear{1997}.
\newblock Finite-state analysis of {SSL 3.0} and related protocols.
\newblock In \bibsc{H.~Orman and C.~Meadows} Eds., \bibemphic{Workshop on
  Design and Formal Verification of Security Protocols} (Sept. 1997). DIMACS.

\bibitem[\protect\citeauthoryear{Paulson}{Paulson}{1994}]{paulson-isa-book}
\bibsc{Paulson, L.~C.} \bibyear{1994}.
\newblock \bibemph{Isabelle: A Generic Theorem Prover}.
\newblock Springer.
\newblock LNCS 828.

\bibitem[\protect\citeauthoryear{Paulson}{Paulson}{}]{paulson-yahalom}
\bibsc{Paulson, L.~C.}
\newblock Relations between secrets: Two formal analyses of the {Yahalom}
  protocol.
\newblock \bibemphic{Journal of Computer Security}.
\newblock in press.

\bibitem[\protect\citeauthoryear{Paulson}{Paulson}{1998}]{paulson-jcs}
\bibsc{Paulson, L.~C.} \bibyear{1998}.
\newblock The inductive approach to verifying cryptographic protocols.
\newblock \bibemphic{Journal of Computer Security}~\bibemph{6}, 85--128.

\bibitem[\protect\citeauthoryear{Ryan and Schneider}{Ryan and
  Schneider}{1998}]{ryan-attack}
\bibsc{Ryan, P. Y.~A. and Schneider, S.~A.} \bibyear{1998}.
\newblock An attack on a recursive authentication protocol: A cautionary tale.
\newblock \bibemphic{Information Processing Letters}~\bibemph{65},~1 (Jan.),
  7--10.

\bibitem[\protect\citeauthoryear{Wagner and Schneier}{Wagner and
  Schneier}{1996}]{ws-ssl}
\bibsc{Wagner, D. and Schneier, B.} \bibyear{1996}.
\newblock Analysis of the {SSL} 3.0 protocol.
\newblock In \bibsc{D.~Tygar} Ed., \bibemphic{USENIX Workshop on Electronic
  Commerce} (1996), pp.\  29--40. USENIX Association.

\end{thebibliography}

\egroup
\end{document}